\documentclass[sigconf]{acmart}
\usepackage{amssymb}
\usepackage{amsmath}
\usepackage{bm}
\usepackage{xspace}
\usepackage{algorithm}
\usepackage{algpseudocode}
\usepackage{varwidth}

\usepackage{multirow,bigdelim}
\usepackage{booktabs}
\usepackage{epsfig}
\usepackage{pgf}
\usepackage{graphicx}
\usepackage{hyperref}
\usepackage{xspace}
\usepackage{color}
\usepackage{epstopdf}
\usepackage{subfig}
\usepackage{chngpage}
\usepackage{setspace}
\usepackage{float}
\usepackage{multirow}
\usepackage{makecell}
\usepackage{balance}
\usepackage[font=footnotesize,skip=0pt]{caption}
\usepackage{fancyhdr}

\copyrightyear{2019}
\acmYear{2019} 
\setcopyright{iw3c2w3}
\acmConference[WWW '19]{Proceedings of the 2019 World Wide Web Conference}{May 13--17, 2019}{San Francisco, CA, USA}
\acmBooktitle{Proceedings of the 2019 World Wide Web Conference (WWW '19), May 13--17, 2019, San Francisco, CA, USA}
\acmPrice{}
\acmDOI{10.1145/3308558.3313736}
\acmISBN{978-1-4503-6674-8/19/05}

\DeclareGraphicsExtensions{.pdf,.png}
\newcommand*{\minimize}{\operatornamewithlimits{minimize}}

\newcommand*{\ML}{Movielens\xspace}
\newcommand*{\YM}{Yahoo Music\xspace}
\newcommand*{\FX}{Flixster\xspace}
\newcommand*{\MLS}{ML\xspace}
\newcommand*{\FXS}{FX\xspace}
\newcommand*{\YMS}{YM\xspace}
\newcommand*{\NFS}{NF\xspace}

\newcommand*{\MLTM}{Movielens 20M\xspace}

\newcommand*{\NF}{Netflix\xspace}

\newcommand*{\LLORMA}{LLORMA\xspace}
\newcommand*{\MF}{Matrix Factorization\xspace}
\newcommand*{\MFS}{MF\xspace}
\newcommand*{\TMF}{Truncated Matrix Factorization\xspace}
\newcommand*{\TMFS}{TMF\xspace}
\newcommand*{\TMFD}{Truncated Matrix Factorization with Dropout\xspace}
\newcommand*{\TMFDS}{TMF + Dropout\xspace}
\newcommand*{\IFWMF}{Inverse Frequency Weighted Matrix Factorization\xspace}
\newcommand*{\IFWMFS}{IFWMF\xspace}
\newcommand*{\FARP}{Frequency Adaptive Rating Prediction\xspace}
\newcommand*{\FARPS}{FARP\xspace}

\begin{document}

\title{Adaptive Matrix Completion for the Users and the Items in Tail}\thanks{This work was supported in part by NSF (1447788, 1704074, 1757916, 1834251), Army
Research Office (W911NF1810344), Intel Corp, and the Digital Technology Center at the
University of Minnesota. Access to research and computing facilities was provided by
the Digital Technology Center and the Minnesota Supercomputing Institute.}

\author{Mohit Sharma}
\affiliation{%
  \institution{University of Minnesota}
  \city{Twin Cities}
  \state{MN}
  \country{USA}
}
\email{sharm163@umn.edu}

\author{George Karypis}
\affiliation{%
  \institution{University of Minnesota}
  \city{Twin Cities}
  \state{MN}
  \country{USA}
}
\email{karypis@umn.edu}

\begin{abstract} %\small\baselineskip=10pt 
Recommender systems are widely used to recommend the most appealing
items to users. These recommendations can be generated by applying
collaborative filtering methods. The low-rank matrix completion method is the 
state-of-the-art collaborative filtering method.  
In this work, we show that the skewed distribution of ratings in the
user-item rating matrix of real-world datasets affects the accuracy 
of  matrix-completion-based approaches. 
Also, we show that the number of ratings that an item or a user has positively
correlates with the ability of low-rank matrix-completion-based approaches to predict the
ratings for the item or the user accurately. 
Furthermore, we use these insights to develop four matrix
completion-based approaches, i.e., \FARP (\FARPS), \TMF (\TMFS), \TMFD (\TMFDS) and \IFWMF (\IFWMFS), that outperforms traditional matrix-completion-based approaches for the users and the items with few ratings in the user-item rating matrix.

\end{abstract}




\maketitle

\section{Introduction} \label{intro}
Recommender systems are used in e-commerce, social networks, and web
search to suggest the most relevant items to each user. Recommender systems
commonly use methods based on Collaborative Filtering~\cite{SarwarKarypis01},
which rely on 
historical preferences of  users over items in order to generate
recommendations. 
These methods predict the ratings for the items not rated by
the user and then select the unrated items with the highest predicted ratings as 
recommendations to the user.

In practice, a user may not rate all the available items,  and hence we observe only a subset of the user-item rating
matrix. 
For the task of recommendations, we need to complete the matrix by predicting the missing
ratings and select the unrated items with high predicted ratings as
recommendations for a user.
The matrix completion approach~\cite{CandesRecht09} assumes that the user-item rating matrix is low
rank and estimates the missing ratings based on the observed ratings in the matrix.
The state-of-the-art  collaborative filtering
methods, e.g., \MF (\MFS)~\cite{Koren2009} are based on the matrix completion
approach.

Assuming that the user-item rating matrix is low-rank, it was
shown that in order to accurately recover the underlying low-rank model of a $n
\times n$ matrix of rank $r$, at least $O(nr \log(n))$ entries in the matrix
should be sampled uniformly at random~\cite{CandesTao2010}.
However, most real-world rating matrices exhibit a skewed distribution of
ratings as some users have provided ratings to few items and certain items have
received few ratings from the users. This skewed distribution may result in
insufficient ratings for certain users and items, and can negatively affect the
accuracy of the matrix completion-based methods. 

This paper investigates how the skewed distribution of ratings in the user-item rating matrix affects the accuracy of the matrix completion-based methods and shows by extensive experiments on different low-rank synthetic datasets and as well as on real datasets that the matrix completion-based methods tend to have poor accuracy for the items and the users with few ratings.
Moreover, this work illustrates that as we increase the number of latent dimensions, the prediction performance for the items and the users with sufficiently many ratings continues to improve, whereas the accuracy of the items and the users with few ratings degrades. This suggests that because of over-fitting, the matrix completion-based methods for large number of latent dimensions do not generalize well for the items and the users with few ratings. 

Building on this finding, we develop four matrix completion-based approaches that explicitly consider the number of
ratings received by an item or provided by a user to estimate the rating of the user on the item. Specifically, we introduce (i) \FARP (\FARPS) method, which uses multiple low-rank models for different frequency of the users and the items; (ii) \TMF (\TMFS) method, which estimates a single low-rank model that adapts with the number of ratings a user and an item has; (iii) \TMFD (\TMFDS) method, which is similar to \TMFS but probabilistically select the ranks for the users and the items; and (iv) \IFWMF (\IFWMFS) method, which weighs the infrequent users and items higher during low-rank model estimation.
Extensive experiments on various datasets demonstrate the effectiveness of the proposed approaches over traditional \MFS-based methods by improving the accuracy for the items (up to 53\% improvement in RMSE) and the users (up to 8\% improvement in RMSE) with few ratings.

\iffalse
The rest of the paper is organized as follows. Section~\ref{notation} presents
the notation used in the paper. Section~\ref{related_work} describes the relevant
prior work. Section~\ref{hypoexp} investigates the effect of skewed distribution of ratings on accuracy 
of the matrix completion-based methods. Section~\ref{methods} presents the
proposed approaches. Section~\ref{exp_eval} describes the experimental
evaluation of the proposed approaches. Section~\ref{exp_res} presents the results
of the proposed approach on real datasets.
Finally, Section~\ref{conclusion} discusses the implications of this work and provides some concluding remarks.
\fi

\section{Related Work} \label{related_work}
%!TEX root = paper.tex
The current state-of-the-art methods for rating prediction are based on matrix
completion, and most of them involve factorizing the user-item rating
matrix~\cite{Koren2009, koren2008factorization, hu2008collaborative}.
In this work, our focus is on analyzing the performance of the matrix completion-based \MFS approach and use the derived insights to
develop an approach that performs better for the users and the items with few
ratings in the user-item rating matrix. 
These approaches estimate user-item rating matrix as a product of two
low-rank matrices known as the user and the item latent factors. 
If for a user $u$, the vector $\bm{p}_u \in \mathbb{R}^r$ denotes the $r$ dimensional
user's latent factor and similarly for the item $i$, the vector 
$\bm{q}_i \in \mathbb{R}^r$ represents the $r$ dimensional item's latent factor, then the
predicted rating ($\hat{r}_{u,i}$) for  user $u$ on  item $i$ is given by
\begin{equation} \label{mf_eq}
  \begin{split}
    \hat{r}_{u,i} = \bm{p}_u\bm{q}_i^T.
  \end{split}
\end{equation}
\noindent The user and the item latent factors are estimated by minimizing a regularized
square loss between the actual and predicted ratings 
\begin{equation} \label{mf_obj}
  \begin{aligned}
    \minimize_{\bm{p}_u, \bm{q}_i} & &\frac{1}{2}\sum_{r_{ui} \in R}
    \left(r_{ui} - \bm{p}_u\bm{q}_i^T \right)^2+ \frac{\beta}{2}
    \left(||\bm{p}_u||_2^2 + ||\bm{q}_i||_2^2 \right),
    %\minimize_{\bm{p}_u, \bm{q}_i} & &\frac{1}{2}\sum_{r_{ui} \in R}
    %\left(r_{ui} - \bm{p}_u\bm{q}_i^T \right)^2+ \frac{\beta}{2}
    %\left(\sum_{u = 1}^{|U|}||\bm{p}_u||_2^2 + \sum_{i = 1}^{|I|}||\bm{q}_i||_2^2 \right),
  \end{aligned}
\end{equation}
\noindent where $R$ is the user-item rating matrix, $r_{ui}$ is the observed rating of user $u$ on item $i$, and  parameter $\beta$ controls the Frobenius norm regularization of the latent factors to prevent overfitting.
%where $R$ is the user-item rating matrix, $r_{ui}$ is the observed rating of user $u$ on item $i$, $|U|$ is the number of users, $|I|$ is the number of items and the parameter $\beta$ controls the Frobenius norm regularization of the latent factors to prevent overfitting.

In another related work~\cite{xin2017folding}, it was shown that the lack of uniform distribution of ratings in the observed data could lead to folding, i.e., the unintentional affinity of dissimilar users and items in the low-rank space estimated by matrix completion-based methods. For example, the absence of an explicit rating between a child viewer and a horror movie could lead to an estimation of the corresponding latent factors such that both the child viewer and the horror movie are close in the low-rank space leading to the erratic recommendation of horror movies to the child viewer. We believe that the non-uniform distribution of the ratings is the reason behind this phenomenon.

The non-uniform distribution of ratings can also be viewed as an instance of rating data \emph{missing not at random} (MNAR)~\cite{little2014statistical}. The proposed solutions to MNAR model the missing data to improve the generated recommendations~\cite{Steck2010MNAR, Steck2013EvalRec,schnabel2016recommendations,marlin2009collaborative}.
However, in our work, we focus on the skewed distribution of ratings which often comes as a result of either new items or new users that are added to the system, or the items that are not popular to get many ratings. We use our analysis
to develop a matrix completion-based approach to improve rating prediction for the
users who have provided few ratings on items or for the items that have
received few ratings from the users.  
  
\iffalse
The recommendations of the users and the items with few ratings may also be improved by applying cold-start recommendation methods~\cite{agarwal2009regression, sharma2015feature, elbadrawy2015user}. These methods rely on the characteristics of the users or the items to generate recommendations. However, we may not always have access to the user or the item characteristics in a recommender system, and we have to rely on the feedback associated with them to generate recommendations. In this work, we focus on improving recommendations for the users and the items with few ratings when we do not have access to their attributes.
\fi

%!TEX root = paper.tex

\section {Impact of Skewed Distribution} \label{hypoexp}

As described in Section~\ref{intro}, the matrix completion-based methods
can accurately recover the underlying low-rank model of a given low-rank matrix
provided entries are observed uniformly at random from the matrix.
However, the ratings in the user-item rating matrix in real-world datasets
represent a skewed distribution of entries because 
some users have provided ratings to few items and certain items have received 
few ratings from the users.

%\subsection{Experiment design} \label{sec:expdesign}

In order to study how the skewed distribution of ratings in real datasets
affects the ability of matrix completion to accurately complete the matrix (i.e.,
predict the missing entries) we performed a series of experiments using synthetically
generated low-rank rating matrices. 
In order to generate a rating matrix $R\in \mathbb{R}^{n\times m}$ of rank $r$ we
followed the following protocol. We started by generating two matrices
$A\in\mathbb{R}^{n\times r}$ and $B\in\mathbb{R}^{m\times r}$ whose values are
uniformly distributed at random in $[0, 1]$. We then computed the singular value
decomposition of these matrices to obtain $A=U_A\Sigma_A V_A^T$ and $B=U_B\Sigma_B
V_B^T$. We then let $P=\alpha U_A$ and $Q=\alpha U_B$ and $R = PQ^T$. Thus, the final
 matrix $R$ of rank $r$ is obtained as the product of two randomly generated rank $r$
matrices whose columns are orthogonal. The parameter $\alpha$ was
determined empirically in order to produce ratings in the range of $[-10, 10]$.

We used the above approach to generate full rating matrices whose dimensions are
those of the two real-world datasets, i.e., \FX (\FXS) and \ML (\MLS),  shown in Table~\ref{table:datasets_table}. For
each of these matrices we select the entries that correspond to the actual user-item pairs that are present in
the corresponding dataset and give it as input to the matrix completion algorithm.
For each dataset we generated five different sets of matrices using different random
seeds and we performed a series of experiments using synthetically generated low-rank
matrices of rank 5 and 20. For each rank, we report the average of performance
metrics in each set from the estimated low-rank models over all the synthetic
matrices.

%!TEX root = paper.tex

\subsection{Results} \label{results}
\begin{table}[bt]\footnotesize
  \centering
  \caption{Datasets used in experiments} \label{table:datasets_table}
    \parbox[t]{8cm}{
    \begin{tabular}{
        @{\hspace{2pt}}l@{\hspace{3pt}}|
        @{\hspace{2pt}}r@{\hspace{3pt}}
        @{\hspace{2pt}}r@{\hspace{3pt}}
        @{\hspace{2pt}}r@{\hspace{3pt}}
        @{\hspace{2pt}}r@{\hspace{3pt}}
        @{\hspace{2pt}}r@{\hspace{3pt}}
        @{\hspace{2pt}}r@{\hspace{3pt}}
        @{\hspace{2pt}}r@{\hspace{3pt}}
        @{\hspace{2pt}}r@{\hspace{3pt}}
        @{\hspace{2pt}}r@{\hspace{3pt}}
      }
      %\toprule
      \hline
      \textbf{Dataset} & \textbf{users} & \textbf{items}  & \textbf{ratings} &
           \textbf{${\mu_u}^{a}$} &
            \textbf{${\sigma_u}^{b}$} &
            \textbf{${\mu_i}^{a}$} & 
            \textbf{${\sigma_i}^{b}$} &
            \%\textsuperscript{\textdagger} \\
            %\textbf{density    (\%)}\tnote{\textdagger} \\
      \hline
    \FX (\FXS) & 147K & 48K & 8.1M &  55 & 226 & 168 & 934 & 1e-1 \\
    \ML (\MLS) & 229K & 26K & 21M  & 92 & 190 & 786 & 3269 & 3e-1 \\
    \YM (\YMS) & 143K & 136K & 9.9M & 69 & 199 & 73 & 141 & 5e-4 \\
    \NF (\NFS) & 354K & 17K & 9.5M &  27 & 59 & 535 & 1693 & 1e-3 \\
         
      \hline
    \end{tabular}

    \begin{flushleft}
      \textsuperscript{a} Average ratings per user ($\mu_u$) or per item ($\mu_i$).\\
      \textsuperscript{b} Standard deviation of ratings per user ($\sigma_u$) or per item ($\sigma_i$).\\
      \textsuperscript{\textdagger} The percentage of observed ratings in the dataset.
    \end{flushleft}
    }
  
  \vspace{-2.25em}
\end{table}

\subsubsection{Effect of item frequency in synthetic datasets} \label{freqeff}
In order to investigate if the number of ratings an item has, i.e., item frequency, has any
influence on the accuracy of the matrix completion-based methods for the item,
we ordered all the items in decreasing order by their frequency in the
rating matrix. Furthermore, we divided these ordered items into ten buckets and for a
user computed the RMSE for items in each bucket based on the error between the
predicted rating by the estimated low-rank model and the ground-truth rating. We repeated this for all the
users and computed the average of the RMSE of the items in each bucket over all the users. 

\begin{figure}[hbt]
  %\centering
  %\hspace*{-1cm}
  \includegraphics[scale=0.6]{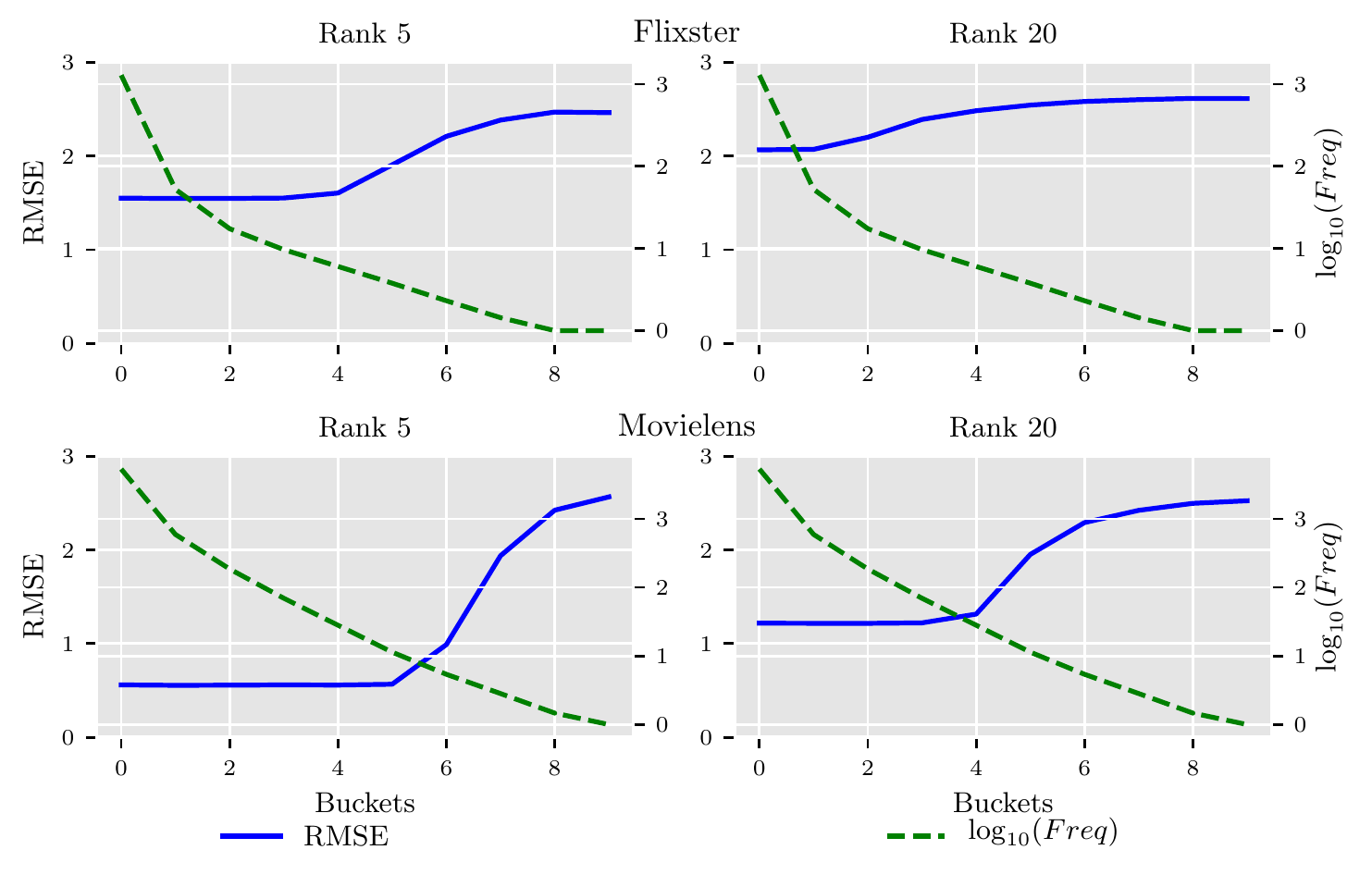}
  \caption{RMSE of the predicted ratings as the frequency of the items decreases.}
  \label{fig:freqRMSEbuck}
  \vspace{-1.5em}
\end{figure}

Figure~\ref{fig:freqRMSEbuck} shows the RMSEs across the buckets along with the
average frequency of the items in the buckets.
As can be seen in the figure, the predicted ratings
for the frequent items tend to have lower RMSE in contrast to infrequent items
for all the datasets. 

\begin{figure}[hbt]
  \centering
  \includegraphics[scale=0.60]{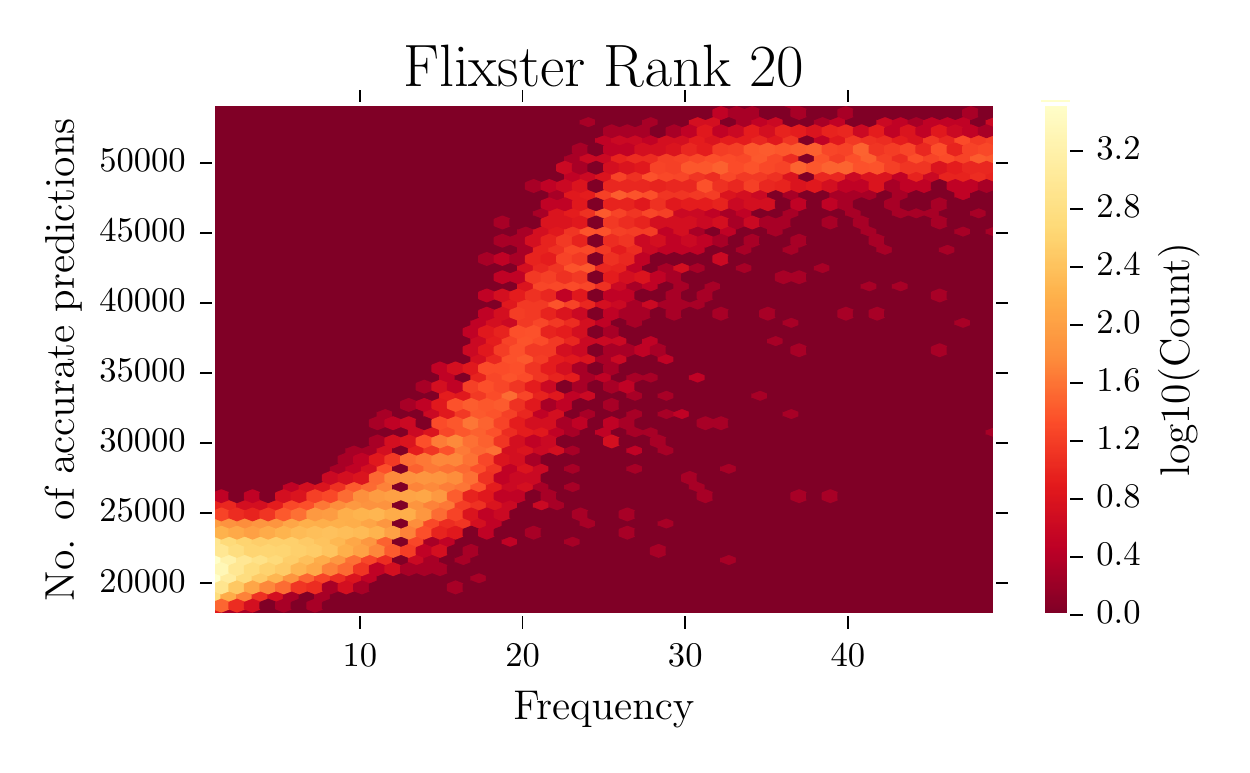} 
  \caption{Scatter map of items having different frequency against their number of
  accurate predictions (Mean absolute error (MAE) $\le$ 0.5) for low-rank
model with rank 20.}
  \label{fig:freq_accu}
  \vspace{-1.5em}
\end{figure}

\begin{figure}[hbt]
  \includegraphics[scale=0.56]{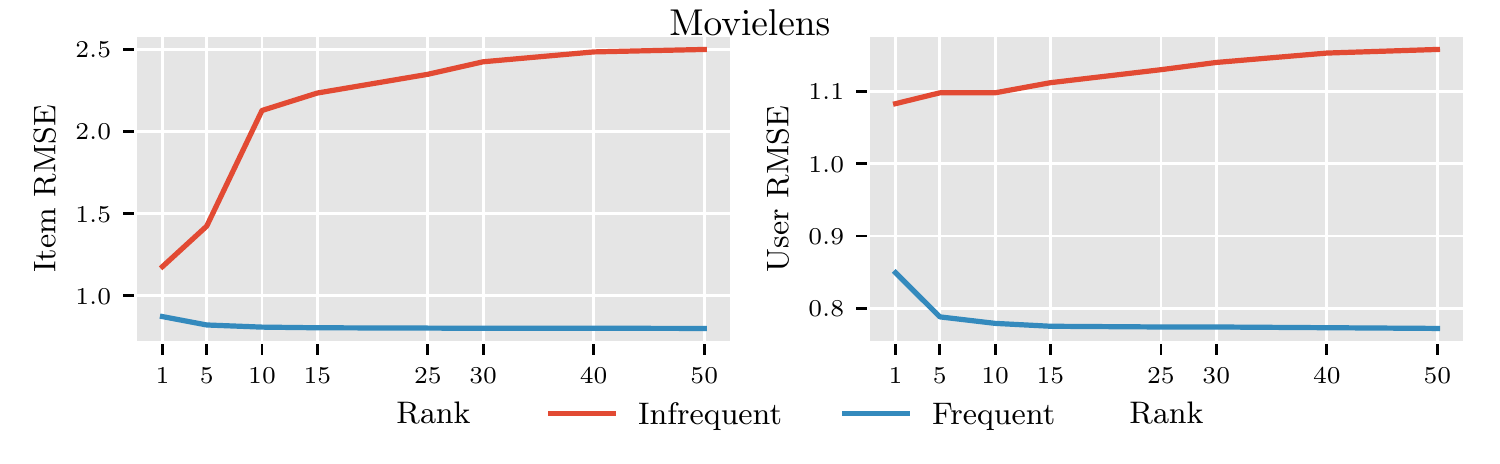}
  \caption{Variation in Test RMSE with increase in rank for the items and the users with different frequency.}
  \label{fig:freq_inFreq_rmse}
  \vspace{-1.5em}
\end{figure}

Figure~\ref{fig:freq_accu} shows the scatter map of items in \FXS having different frequency
against the number of instances where the absolute difference between the
original and the predicted rating, i.e., \emph{Mean Absolute Error (MAE)},  is
$\le 0.5$. 
As can be seen in the figure, the number of accurate predictions is
significantly lower for items having fewer ratings ($\le 20$) compared to that of
the items having a large number of ratings ($\ge 30$). 
The lower error of the frequent items is
because they have sufficient ratings to estimate their latent
factors accurately. 
Hence for the real datasets, items appearing at the top in ordering by
frequency and having high predicted scores will form a reliable set of recommendations to a user.

\subsubsection{Effect of frequency  on accuracy in real datasets}
In order to assess the finding that the infrequent items are not
estimated accurately by the matrix completion method, we evaluated matrix
completion on a random held-out subset of the real datasets. We followed the standard procedure of  cross-validation and exhaustive grid search for hyperparameters for model selection.  
We computed RMSE over the infrequent items in the test split, i.e., the items that
have few ratings in the training split. 
For the analysis, we ordered the items in increasing order by the number of
ratings in training splits. Next, we divided these ordered items into quartiles
and identified the items in the first and the last quartile as the infrequent
and the frequent items, respectively.

Figures~\ref{fig:freq_inFreq_rmse} show 
the RMSE for the items and the users in the test for the \ML (\MLS) dataset.
We can see that the RMSE of the frequent
items (or users) is lower than that of the infrequent items (or users). Furthermore, we observed similar trends in the remaining datasets (results not shown here due to space constraints).  
These results suggest that the matrix completion method fails to estimate the
preferences for the infrequent items (or users) accurately in the real datasets.
Also, the RMSE of the 
infrequent items  increases with the increase in the
rank while that of frequent items decreases with the increase in the rank. Similarly, the RMSE of the infrequent users 
increases with the increase in the rank.
The increase in RMSE with the increase in ranks suggests that infrequent items or infrequent users may not 
have sufficient ratings to estimate all the ranks accurately thereby leading to the error in
predictions for such users or items. The finding that infrequent items or infrequent users have better accuracy for fewer ranks follows from the result that $O(nr \log(n))$ entries are required to recover the underlying low-rank model of a $n \times n$ matrix of rank $r$~\cite{CandesTao2010}, and therefore for fewer entries (e.g., infrequent users or infrequent items) we may recover only fewer ranks of the underlying low-rank model accurately.

\section{Methods}\label{methods} 
%!TEX root = paper.tex

The analysis presented in the previous section showed that as the underlying rank of the low-rank model that describes the data increases, the error associated with estimating such a low-rank model from the skewed data increases for the infrequent users and the infrequent items.  
We use these observations to devise multiple approaches to improve the accuracy of the low-rank models for such users and items. 

\subsection{\FARP (\FARPS)}\label{farp}
Since the error of the predictions from the estimated low-rank models increases for the infrequent users or items in skewed data, we propose to estimate lower dimensional latent factors for the infrequent users or items, and estimate higher dimensional latent factors for the frequent users or items.
In this approach, we propose to learn multiple low-rank models with different ranks from all the available data and while predicting the rating of a user on an item we select the model that performed the best for the infrequent user or the item associated with the rating. Hence, the predicted rating of user $u$ on item $i$ is given by
\begin{equation}
    \hat{r}_{ui} = \bm{p}_{uk}\bm{q}_{ik}^T,
\end{equation}
\noindent where $\bm{p}_{uk}$ and $\bm{q}_{ik}$ are the user and the item latent factors from the $k$th low-rank model. For example, if $f_u < f_i$ then we select the $k$th low-rank model for prediction such that it has the best performance for users having frequency $f_u$, and similarly if $f_i < f_u$ then we select the model with the best performance for items with frequency $f_i$.

The user and items can be assigned to different low rank models based on the number of ratings that exists for them in the dataset. One approach that we investigated is to order the users and the items by the number of ratings and divide them into equal quartiles and save the best performing model for each quartile.

\subsection{\TMF (\TMFS)}
An alternate approach we develop is to estimate only a subset of the ranks 
for these users or items. In this approach, the estimated rating for user $u$ on item $i$ is given by
\begin{equation} \label{mf_drop_est_eq} 
  \begin{split}
  \hat{r}_{u,i} = \bm{p_u}(\bm{q_i}\odot\bm{h_{u,i}})^T,
  \end{split}
\end{equation}
where $\bm{p_u}$ denotes the latent factor of user $u$, $\bm{q_i}$ represents the
latent factor of item $i$,  $\bm{h_{u,i}}$ is a vector containing $1$s in the
beginning followed by $0$s, and 
$\odot$ represents the elementwise Hadamard product between the vectors.
The vector $\bm{h_{u,i}}$ is used to select the ranks that are \emph{active} for the
$(u,i)$ tuple. The $1$s in $\bm{h_{u,i}}$ denote the active ranks for the
$(u,i)$ tuple.

\subsubsection{Frequency adaptive truncation}
One approach that we investigated for selecting the active ranks, i.e., $\bm{h_{u,i}}$, for a user-item rating is
based on the frequency of the user and the item in the rating matrix. 
%If either
%the user or the item is infrequent then depending on the frequency the weight is
%0 for later ranks as we may not have sufficient ratings to
%estimate the later ranks accurately.  
In this approach, for a given rating by a user on an item, first, we determine
the number of ranks to be updated based on either the user or the item depending
on the one having a lower number of ratings. 
In order to select the ranks, we normalize the frequency of the user and the
item,
and use a non-linear activation function, e.g., sigmoid function, to map this frequency of the user
or the item in [0, 1]. Finally, we used this mapped value as the number of
active ranks 
selected for the update of the user and the item latent factors. The number of
active ranks to be selected is given by
\begin{equation}
  k_{u,i} = 
    \frac{r}{1 + e^{-k(f_{min} -  z)}},
\end{equation}
\noindent where $r$ is the dimension of the user and the item latent factors, $f_{min} = \min(f_u, f_i)$, $f_u$ is the frequency of user $u$, $f_i$ is the frequency of item $i$, $k$
controls the steepness of the sigmoid function and $z$ is the value
of the sigmoid's midpoint. 
%An additional motivation for using a non-linear activation function, e.g.,
%sigmoid function, is that the shape of the plot in Figure~\ref{fig:freq_accu} is non-linear.
The use of such a function assists in identifying the users or the items that
can not be estimated accurately using all the ranks and we can only estimate
few ranks more accurately for such users or items. The active ranks for a user or an item can be chosen either from the beginning of all ranks or end of all ranks or can be chosen arbitrarily among all ranks until the same active ranks are used consistently for the user and the item. For ease of discussion and simplicity, we will assume that active ranks are chosen from the beginning of all ranks. 
Hence, the active ranks to be selected are given by

\begin{equation} \label{rank_wt_eq}
  \bm{h_{u,i}}[j] = 
  \begin{cases}
    1,& \text{if } j \leq k_{u,i} \\
    0,& \text{otherwise}.
  \end{cases}
\end{equation}

We will refer to this method as Truncated Matrix Factorization (TMF).

\subsubsection{Frequency adaptive probabilistic truncation}

An alternative way to select the active ranks is to assume that the
number of active ranks follows a Poisson distribution with parameter
$k_{u,i}$. 
This method is similar to Dropout~\cite{srivastava2014dropout} technique in neural networks, where parameters
are selected probabilistically for updates during learning of the model.
Similar to regularization it provides a way of preventing overfitting in
learning of the model. 
The active ranks to be selected  are given by
\[
  \bm{h_{u,i}}[j] = 
  \begin{cases}
    1,& \text{if } j \leq \theta_{u,i} \\
    0,& \text{otherwise},
  \end{cases}
\]
\noindent where $\theta_{u,i} \sim Poisson(k_{u,i})$.
We will call this method as \TMFD (\TMFDS). Similar to Equation~\ref{mf_obj}, the parameters of the model, i.e., the user and the item latent factors can be estimated by minimizing a regularized
square loss between the actual and the predicted ratings.

\subsubsection{Rating prediction}
After learning the model the predicted rating for user $u$ on item $i$ for \TMFS model is given by
\begin{equation}
  \begin{split}
    \hat{r}_{u,i} = \bm{p_u}(\bm{q_i}\odot\bm{h}_{u,i})^T,
  \end{split}
\end{equation}
\noindent where the active ranks, i.e., $\bm{h_{u,i}}$, is given by
Equation~\ref{rank_wt_eq}. The predicted rating for the user and the item under \TMFDS model is given
by the least number of ranks for whom the cumulative distribution function (CDF)  for
Poisson distribution with parameter $k_{ui}$ obtains approximately the value of
1. The active ranks, i.e., $\bm{h_{u,i}}$, for prediction under \TMFDS
are given by
\[
  \bm{h_{u,i}[j]} = 
  \begin{cases}
    1,& \text{if } j \leq s \\
    0,& \text{otherwise},
  \end{cases}
\]
\noindent where $s$ is the least number of ranks for whom the CDF, i.e, $P(x <= s) \approx 1
$ and $x \sim Poisson(k_{u,i})$.

Unlike \FARPS, which requires us to estimate multiple models, \TMFS estimates a single model however it involves tuning of more hyperparameters in comparison to FARP.

\subsection{\IFWMF (\IFWMFS)}
In addition to the above approaches, we explored a weighted matrix factorization-based approach where we weigh the reconstruction error higher for the infrequent users and items. 
We propose to estimate the user and the item latent factors by minimizing a regularized weighted square error loss between the actual and the predicted ratings
\begin{equation} \label{mf_wt_obj}
  \begin{aligned}
    \minimize_{\bm{p_u}, \bm{q_i}} & &\frac{1}{2}\sum_{r_{ui} \in R}
    w_{ui}\left(r_{ui} - \bm{p}_u\bm{q}_i^T \right)^2+ \frac{\beta}{2}
    \left(||\bm{p}_u||_2^2 + ||\bm{q}_i||_2^2 \right),
  \end{aligned}
\end{equation}
\noindent where the the weight $w_{ui}$ is given by
\begin{equation} \label{inv_wt_eq}
     w_{ui} = \frac{1}{1.0 + \rho f_{\min}},
\end{equation}
\noindent where $\rho$ is a constant, $f_{\min} = \min(f_u, f_i)$, $f_u$ and $f_i$ are the normalized frequency of user $u$ and item $i$, respectively. Essentially, we weigh the error in predictions more for the infrequent users and the infrequent items. 
This resembles the weighted matrix factorization~\cite{hu2008collaborative,koenigstein2013towards} where the weight of the reconstruction error is proportional to the confidence on the observed rating of user $u$ on item $ i$ however in our method we weigh the error inversely proportional to the frequency of ratings observed for user $u$ or item $i$.
This is similar to the inverse propensity model-based approach~\cite{schnabel2016recommendations}, where the propensity is proportional to the frequency of the user or the item. The up-weighting of the reconstruction error associated with the infrequent users or the infrequent items may lead to over-fitting as we only have few ratings for these users and items.

\section{Experimental Evaluation}\label{exp_eval}
%!TEX root = paper.tex

\subsection{Comparison algorithms}\label{subsection:comparison_algos}

We compared our proposed approaches\footnote{ Code for reference is available at
\href{https://github.com/mohit-shrma/matfac}{github.com/mohit-shrma/matfac}.} against the the state-of-the-art \MF~\cite{koren2008factorization} and \LLORMA~\cite{lee2013local} method.
\LLORMA assumes that the different parts of the user-item rating matrix can be approximated by different low-rank models and the complete user-item rating matrix is approximated as a weighted sum of these individual low-rank models.
We have used the LibRec~\cite{guo2015librec} software package to compare the proposed methods against the \LLORMA approach.

\subsection{Model selection}
We performed grid search to tune the dimensions of the latent factors,
regularization hyper-parameters, constant ($\rho$), and sigmoid function's parameters, i.e., $k$ and
$z$. We searched for regularization weights ($\lambda$) in the range [0.001,
0.01, 0.1, 1, 10], dimension of latent factors ($r$) in the range [1, 5, 10, 15,
25, 50, 75, 100], constant ($\rho$) in the range [1, 10, 50],  steepness constant ($k$) in the range [1 5, 10, 20, 40], and mid-point
($z$) in the range [-0.75, -0.50, -0.25, 0, 0.25, 0.50, 0.75]. The final parameters were selected based on the performance on the validation split. For \LLORMA, we varied the number of local models ($l_m$) in the range [1, 5, 8, 15, 25, 50]. 

For \FARPS, we ordered the users in ascending order by frequency and divided them into equal quartiles. For each quartile, we saved the best performing model, i.e., the model having the lowest RMSE for all the users in that quartile in the validation split. Similarly, we ordered the items in ascending order by frequency and divided them into equal quartiles. Similar to users, we saved the best performing model for each quartile of items. At the time of prediction of rating for a user on an item, 
we choose the the model associated with the quartile of the  user if the user is having lower number of ratings than the item, and if the item is having lower number of ratings than the user than we choose the model associated with the quartile of the item . 

\subsection{Datasets}\label{datasets}
In addition to \FX (\FXS)~\cite{flixsterweb} and \MLTM (\MLS)~\cite{harper2016movielens} datasets, we evaluated our proposed methods on subsets of the \YM (\YMS)~\cite{ymusic,schelter2013all} and \NF (\NFS)~\cite{bennett2007netflix} datasets that we created in order to have a skewed distribution. These datasets were generated as follow. 
First, for each user we randomly selected the number of ratings that we want to sample from the user's ratings and randomly sampled these ratings for all users. Next, from the sampled ratings in previous step, for each item we randomly selected the number of ratings that an item has received from users in sampled ratings and randomly sampled  these ratings for all the items. After following the above two steps, the sampled ratings from these datasets follows a skewed distribution and characteristics of all the datasets used in experiments are presented in Table~\ref{table:datasets_table}.

\subsection{Evaluation methodology}
To evaluate the performance of the proposed methods we divided the available
ratings in different datasets into training, validation and test splits by
randomly selecting 20\% of the ratings for each of the validation and the test
splits. The validation split was used for model selection, and the model that
was selected was used to predict ratings on the test split. We repeated this
process three times and report the average RMSE across the runs.

In addition to computing RMSE obtained by different methods for the ratings in
the test split, we also investigated the performance of the proposed approaches
for the items and the users with a different number of ratings in the training
split. To this end, we ordered the items and the users in increasing order by
their number of ratings in training split and divided them equally into quartiles. We will report the RMSE achieved by different methods for ratings in
the test split for the users and the items in these quartiles.

\section{Results and Discussion}\label{exp_res}
%!TEX root = paper.tex

\begin{table*}[hbt]\footnotesize
\caption{Test RMSE of the proposed approaches for different datasets. The RMSE for the users and the items in different quartiles order by their frequency. Q1 refers to
the quartile containing the least frequent users or items followed by remaining
in Q2, Q3, and Q4. Table~\ref{table:test_ratings_quartiles} shows the average number of test ratings in different quartiles for different datasets.}

\label{table:drop_test_quart_rmse}

    \begin{center}
        \begin{tabular}{l|rrrrrr|rrrrrr}
                 \toprule
                 & \multicolumn{6}{c}{\centering \FX (\FXS)} & \multicolumn{6}{c}{\centering \ML (\MLS)} \\
                 \midrule
                 & \MFS  & \LLORMA & \IFWMFS & \FARPS & \TMFS & \TMFDS & \MFS & \LLORMA
                 & \IFWMFS & \FARPS & \TMFS & \TMFDS \\
                 \midrule
                 No. of low-rank models
                    & 1 & 20 & 1 & 8 & 1 & 1
                    & 1 & 8 & 1 & 8 & 1 & 1 \\
                 \midrule
                 Rank & 15 & 50 & 15 & NA\textsuperscript{\textdagger} & 10 & 15
                    & 100 & 25 & 100 & NA\textsuperscript{\textdagger} & 100 & 25 \\
                 \midrule    
                 All test ratings & 0.864 & 0.871 & 0.867 & 0.863 & 0.851 & \underline{0.847}
                    &  0.806 & 0.806 & 0.804 & \underline{0.797} & 0.804 & 0.804  \\
                 \midrule
                 Item Q1 & 1.302 & 1.705 & 1.289 & 1.258 & \underline{1.252} & 1.256
                    & 2.527 & 1.501 & 2.382 & \underline{1.178} & 2.377 & 2.115 \\
                 Item Q2 & 0.961 & 1.099 & 0.961 & 0.962 & \underline{0.944} & \underline{0.944} 
                    & 1.619 & 0.974 & 1.449 & \underline{0.937} & 1.499 & 1.123 \\
                 Item Q3 & 0.800 & 0.863 & 0.801 & 0.798 & 0.785 & \underline{0.780}
                    & 0.891 & \underline{0.841} & 0.869 & 0.855 & 0.866 & 0.851 \\
                 Item Q4 & 0.864 & 0.865 & 0.867 & 0.863 & 0.851 & \underline{0.847}
                    & 0.799 & 0.805 & 0.799 & \underline{0.795} & 0.798 & 0.801 \\
                 \midrule
                 User Q1 & 1.292 & 1.388 & 1.261 & \underline{1.246} & 1.247 & 1.260
                    &  1.174 & 1.092 & 1.120 & 1.083 & 1.115 & \underline{1.078} \\
                 User Q2 & 1.177 & 1.255  & 1.156 & \underline{1.143} & 1.144 & 1.151
                    & 0.975 & \underline{0.964} & 0.965 & 0.965 & 0.\underline{964} & 0.968 \\
                 User Q3 & 0.974 & 1.002 & 0.969 & 0.967 & \underline{0.964} & \underline{0.964} 
                    & 0.853 & 0.856 & \underline{0.852} & 0.853 & \underline{0.852} & 0.863\\
                 User Q4 & 0.853 & 0.852 & 0.857 & 0.853 & 0.841 & \underline{0.836}
                    & 0.767 & 0.774 & 0.769 & \underline{0.761} & 0.769 & 0.769 \\
                 %\bottomrule
            \end{tabular}
    \end{center}

    \begin{center}
        \parbox[t]{16.7cm}{
            \begin{tabular}{l|rrrrrr|rrrrrrr}
                 \toprule
                 & \multicolumn{6}{c}{\centering \YM (\YMS)\textsuperscript{\textasteriskcentered} } & \multicolumn{6}{c}{ \centering \NF (\NFS)\textsuperscript{\textasteriskcentered}} \\
                 \midrule
                 & \MFS & \LLORMA & \IFWMFS & \FARPS & \TMFS & \TMFDS & \MFS & \LLORMA & \IFWMFS & \FARPS & \TMFS & \TMFDS \\
                 \midrule
                 No. of low-rank models 
                    & 1 & 25 & 1 & 8 & 1 & 1
                    & 1 & 25 & 1 & 8 & 1 & 1 \\
                 \midrule
                 Rank & 100 & 5 & 75 & NA\textsuperscript{\textdagger} & 75 & 75
                    & 75 & 25 & 75 & NA\textsuperscript{\textdagger} & 75 & 75 \\
                 \midrule
                 All test ratings & 1.170 & 1.177 & 1.162 & \underline{1.152} & 1.163 & 1.164
                    & 0.906 & 0.903 & 0.901 & \underline{0.901} & 0.903 & 0.904  \\
                 \midrule
                 Item Q1 & 1.245 & 1.250 & 1.238 & 1.235 & 1.239 & \underline{1.234}
                    & 1.067 & 1.059 & 1.054 & \underline{1.047}  & 1.051 & 1.060 \\
               
                 Item Q2 & 1.165 & 1.171 & 1.159 & \underline{1.152} & 1.161 & 1.162
                    & 1.018  & 1.006 & 1.007  & \underline{1.005}  & 1.008 & 1.014 \\
               
                 Item Q3 & 1.154 & 1.164  & 1.149 & \underline{1.139} & 1.151 & 1.152
                    & 0.967 & 0.964 & 0.956 & \underline{0.955}  & 0.959 & 0.962 \\
               
                 Item Q4 & 1.170 & 1.172 & 1.161 & \underline{1.150} & 1.162 & 1.163
                    & 0.896 & 0.893 & \underline{0.891} & 0.892  & 0.893 & 0.894 \\
               
                 \midrule
               
                 User Q2 & 1.429 & 1.452 & 1.412 & 1.408 & 1.407 & \underline{1.395}
                    & 1.281 & 1.378 & \underline{1.258} & 1.274 & 1.259 & 1.256  \\
               
                 User Q3 & 1.236 & 1.231 & 1.220 & \underline{1.210} & 1.224 & 1.225
                    & 1.018 & 1.011 & \underline{1.008} & 1.006 & 1.010 & 1.013 \\
               
                 User Q4 & 1.155 & 1.160 & 1.148 & \underline{1.138} & 1.150 & 1.15
                    & 0.879 & \underline{0.874} & 0.875 & 0.875 & 0.877 & 0.877  \\
                 \bottomrule
            \end{tabular}
            \begin{flushleft}
                \textsuperscript{\textdagger} The ranks used for \FARPS are in Table~\ref{table:ranks_ifwmf}.\\
                \textsuperscript{\textasteriskcentered} Due to sampling, the test splits are not having any ratings for users in Q1 for these datasets.
            \end{flushleft}
        }
    \end{center}
    
    \vspace{-1.5em}
\end{table*}

\subsection{Performance for rating prediction on entire dataset}
Table~\ref{table:drop_test_quart_rmse} shows the results achieved by the proposed methods on the various datasets. As can be seen in the table for the task of rating predictions for all the ratings in the test splits the proposed approaches perform better than the \MFS method for \FXS, \MLS, \YMS and  \NFS datasets. Interestingly, the proposed approaches have performed even better than the state-of-the-art \LLORMA method and this suggests that \LLORMA can be further improved by estimating local low-rank models that considers the skewed distribution of ratings in datasets. We found the difference between the predictions of different methods to be statistically significant  ($p$-value $\le$ 0.01 using two sample $t$-test). The performance is significantly better for \FXS in comparison to that of other datasets.  Additionally, on \FXS dataset, the \MFS method outperforms the \LLORMA method and a possible reason for this is that because of the skewed distribution \LLORMA is not able to estimate a model that is as accurate as a global \MFS model.  
Moreover, \LLORMA needs a significantly large number of local low-rank models in comparison to the proposed approaches. 
Also, by comparing the number of latent dimensions used by the models shown in Table~\ref{table:drop_test_quart_rmse} we can see that, for \MLS, \TMFDS and \LLORMA needs significantly fewer ranks, i.e., 25, in comparison to that of \MFS, i.e., 100, to achieve the same performance and we believe that this could be because of \MFS overfitting the \MLS dataset for higher dimension of latent factors.

\begin{table}[hbt]\footnotesize
    \caption{Rank used by FARP for different datasets.}
    \label{table:ranks_ifwmf}
    \begin{center}
        
            \begin{tabular}{l|rrrr| rrrr}
                \toprule
                     & \multicolumn{4}{c}{Item} &  \multicolumn{4}{c}{User}\\
                    & Q1 & Q2 & Q3 & Q4 & Q1 & Q2 & Q3 & Q4 \\
                \midrule
                    \FXS & 1 & 1 & 15 & 15 & 1 & 1 & 1 & 15 \\
                    \MLS & 1 & 5 & 10 & 100 & 1 & 15 & 30 & 100 \\
                    \YMS & 75 & 75 & 100 & 100 & 5 & 5 & 100 & 100 \\
                    \NFS & 50 & 50 & 50 & 75 & 50 & 50 & 50 & 75 \\
                \midrule
            \end{tabular}
        
    \end{center}
    \vspace{-1.5em}
\end{table}

\begin{table}[hbt]\footnotesize
    \caption{Average number of ratings in quartiles of test splits.}
    \centering
    \begin{center}
        \parbox[t]{7.7cm}{
            \begin{tabular}{l|llll| llll}
                \toprule
                     & \multicolumn{4}{c}{Item} &  \multicolumn{4}{c}{User}\\
                    & \makecell{Q1\\($10^3$)} & \makecell{Q2\\($10^3$)} 
                    & \makecell{Q3\\($10^3$)} & \makecell{Q4\\($10^6$)}
                    & \makecell{Q1\\($10^3$)} & \makecell{Q2\\($10^3$)}
                    & \makecell{Q3\\($10^3$)} & \makecell{Q4\\($10^6$)} \\
                \midrule
                    \FXS & 4.8 & 6.3 & 22.0 & 1.4 & 16.5 & 15.9 & 54.7 & 1.4 \\
                    \MLS & 2.7 & 9.5 & 92.5 & 3.7 & 111.2 & 194.8 & 533.5 & 2.9 \\
                    \YMS & 83.4 & 175.4 & 303.2 & 1.2 &  NA\textsuperscript{\textdagger} & 38.9 & 181.4 &  1.6   \\
                    \NFS & 13.6  & 40.1 & 136.0 & 1.5 & 
                    NA\textsuperscript{\textdagger} & 32.0 & 209.8 & 1.4   \\
                \midrule
            \end{tabular}
            \begin{flushleft}
                 \textsuperscript{\textdagger} Due to sampling, the test splits do not have ratings for users in Q1.
            \end{flushleft}
        }
    \end{center}

    \label{table:test_ratings_quartiles}
    \vspace{-1.5em}
\end{table}

\subsection{Performance for the users and the items with different number of
ratings}

Table~\ref{table:drop_test_quart_rmse} also shows the performance achieved by the different methods across the different quartiles of users and items. By comparing the performance of the different schemes we can see that the
proposed methods significantly outperform the \MFS method and state-of-the-art \LLORMA method for lower quartiles for
majority of the datasets. This illustrates the effectiveness of the developed methods for the users and the items with few ratings.
The better performance of the proposed methods for the users and the items with
few ratings is because we can estimate accurately only a few ranks for them,
and unlike \MFS and \LLORMA the proposed approaches are effective in model estimation or predicting the ratings for these users and items.

Among the proposed approaches, the \TMFS-based approaches (\TMFS and \TMFDS), perform better in most of the quartiles in \FXS dataset. Surprisingly, \FARPS consistently performs better than the \MFS and \LLORMA across most of the datasets, and this is promising as compared to \TMFS-based approaches \FARPS has fewer hyperparameters to tune.
Specifically, in \TMFS-based approaches we have to tune regularization weights ($\lambda$), dimension of latent factors ($r$), steepness constant ($k$) and mid-point ($z$), while in 
\FARPS we have to tune only three parameters, i.e., number of low-rank models, regularization weights ($\lambda$) and dimension of latent factors ($r$).
This might be of interest to practitioners because multiple low-rank models under \FARPS can be estimated in parallel using vanilla \MFS widely available in off-the-shelf packages, e.g., SPARK~\cite{ZahariaSpark} and scikit-learn~\cite{scikit-learn}.

While the \IFWMFS method performs better than \MFS in lower quartiles, the other proposed approaches, i.e., \FARPS and \TMFS-based methods, outperform \IFWMFS for most of the quartiles in all the dataset thereby illustrating the effectiveness of \FARPS and \TMFS-based methods in preventing over-fitting and generating better predictions.

\section{Conclusion} \label{conclusion}
%!TEX root = paper.tex

In this work, we have investigated the performance of the matrix
completion-based low-rank models for estimating the missing ratings in real
datasets and its impact on the item recommendations. 
We showed in Section~\ref{hypoexp} that the matrix completion-based methods because
of skewed distribution of ratings fail to predict the missing entries accurately in the matrices thereby leading to an error in
predictions and thus affecting item recommendations. 
Based on these insights we presented different methods in Section~\ref{methods}, which considers the frequency of
both the user and the item to estimate the low-rank model or for predicting the ratings. 
The experiments on real datasets show that the proposed approaches significantly outperforms the
state-of-the-art \MF method for rating predictions for the users and the items having few ratings in the
user-item rating matrix.

\bibliographystyle{ACM-Reference-Format}
\balance
\bibliography{refs}

\end{document}